# Deep Metric Learning-based Image Retrieval System for Chest Radiograph and its Clinical Applications in COVID-19

Aoxiao Zhong [a, b†], Xiang Li [a†], Dufan Wu [a], Hui Ren [a], Kyungsang Kim [a], Younggon Kim [a], Varun Buch [c], Nir Neumark [c], Bernardo Bizzo [c], Won Young Tak [d], Soo Young Park [d], Yu Rim Lee [d], Min Kyu Kang [e], Jung Gil Park [e], Byung Seok Kim [f], Woo Jin Chung [g], Ning Guo [a], Ittai Dayan [b], Mannudeep K. Kalra [a], Quanzheng Li [a, c*]

[a] Department of Radiology, Massachusetts General Hospital, Boston, MA

[b] School of Engineering and Applied Sciences, Harvard University, Boston, MA

[c] MGH & BWH Center for Clinical Data Science, Boston, MA

[d] Department of Internal Medicine, School of Medicine, Kyungpook National University, Daegu, South Korea

[e] Department of Internal Medicine, Yeungnam University College of Medicine, Daegu, South Korea

[f] Department of Internal Medicine, Catholic University of Daegu School of Medicine, Daegu, South Korea

[g] Department of Internal Medicine, Keimyung University School of Medicine, Daegu, South Korea

[†] These authors contribute equally to this work

[*] Corresponding author

## ABSTRACT

In recent years, deep learning-based image analysis methods have been widely applied in computer-aided detection, diagnosis and prognosis, and has shown its value during the public health crisis of the novel coronavirus disease 2019 (COVID-19) pandemic. Chest radiograph (CXR) has been playing a crucial role in COVID-19 patient triaging, diagnosing and monitoring, particularly in the United States. Considering the mixed and unspecific signals in CXR, an image retrieval model of CXR that provides both similar images and associated clinical information can be more clinically meaningful than a direct image diagnostic model. In this work we develop a novel CXR image retrieval model based on deep metric learning. Unlike traditional diagnostic models which aims at learning the direct mapping from images to labels, the proposed model aims at learning the optimized embedding space of images, where images with the same labels and similar contents are pulled together. The proposed model utilizes multi-similarity loss with hard-mining sampling strategy and attention mechanism to learn the optimized embedding space, and provides similar images, the visualizations of disease-related attention maps and useful clinical information to assist clinical decisions. The model is trained and validated on an international multi-site COVID-19 dataset collected from 3 different sources. Experimental results of COVID-19 image retrieval and diagnosis tasks show that the proposed model can serve as a robust solution for CXR analysis and patient management for COVID-19. The model is also tested on its transferability on a different clinical decision support task for COVID-19, where the pre-trained model is applied to extract image features from a new dataset without any further training. The extracted features are then combined with COVID-19 patient's vitals, lab tests and medical histories to predict the possibility of airway intubation in 72 hours, which is strongly associated with patient prognosis, and is crucial for patient care and hospital resource planning. These results demonstrate our deep metric learning based image retrieval model is highly efficient in the CXR retrieval, diagnosis and prognosis, and thus has great clinical value for the treatment and management of COVID-19 patients.

*Keywords:* Chest Radiograph, COVID-19, Image Retrieval, Image Content Query.

# 1. INTRODUCTION

In recent years, thanks to the combined advancement of computational power, accumulated high-quality medical image datasets, and the development of novel deep learning-based artificial intelligence (AI) algorithms, there has been a widespread applications of AI in radiology and clinical practice (Thrall et al., 2018). Various studies have shown superior performance of deep learning methods in extracting low- to high-level image features and learning discriminative representations (i.e. embeddings) from large amount of data (Litjens et al., 2017). As one of the most common imaging modalities for diagnostic radiology exams, chest radiograph (CXR) has been receiving enormous attentions in the field of artificial intelligence-based image analysis because of its importance for public health, wide utilization and relatively low cost (Kallianos et al., 2019). There has been a range of imaging processing studies for CXR using deep learning, including diagnosis of thoracic diseases (Lakhani and Sundaram, 2017; Qin et al., 2018), novel methodology development (Ellis et al., 2020; Pesce et al., 2019), and the establishment of open CXR image databases (Wang et al., 2017c).

## 1.1 Computer-aided diagnostic methods on chest radiograph images in COVID-19

The pandemic of the novel coronavirus disease 2019 (COVID-19) is rapidly spreading throughout the world with a high mortality rate in certain populations. Chest imaging including computer tomography (CT) and CXR has been playing a crucial role in patient triaging, diagnosing and monitoring the disease progression. For instance, when the supply and accuracy of COVID-19 polymerase chain reaction (PCR) testing could not meet the clinical need, chest CT has been recommended as a screening tool in the guideline on COVID-19 management during the early outbreak in Wuhan, China (China, 2020). Particularly, the medical image analysis community has quickly responded by developing novel COVID-19 diagnostic and segmentation solutions, including works by (Wang et al., 2020c) where very high specificity were obtained by a 3D Resnet, works by (Kang et al., 2020) to incorporate multi-view features from CT into diagnosis, works by (Ouyang et al., 2020) to overcome the challenge of imbalanced distribution of lesion regions, works by (Han et al., 2020) which utilizes a generative approach for better scalability and flexibility, works by (Wang et al., 2020a) for simultaneous pneumonia detection and lesion type classification, as well as works by (Fan et al., 2020) for lesion region segmentation. In contrast, under the guideline of American College of Radiology (2020), CT imaging is less commonly used in the U.S. due to the lack of specificity in diagnosis as well as logistic/resource/infection concerns (Hope et al., 2020). On the other hand, chest radiography, especially portable radiography units are considered medically necessary in ambulatory care facilities, since they do not require patient transfer to imaging department and are easier to sterilize. With more evidence on CXR imaging of COVID-19 coming out since January, consistent findings, such as ground glass opacities distributed in both lungs, can be observed and summarized (Ng et al., 2020). These findings suggest the potential of using CXR for severity assessment (based on total lung involvement), monitoring disease progression and predicting patient prognosis. However, it is still challenging even for experienced radiologists to interpret these non-specific findings with confidence, especially on CXR (Choi et al., 2020), since there are a lot of unknown about the novel infectious disease. Therefore, an AI system that can learn from top radiologists and provide consistent results would be very valuable in clinical practice. In response to the shortage of radiologists in handling CXR images especially in developing countries, AI-assisted COVID-19 diagnostic tools have been developed in multiple studies. For example, the CAD4COVID-XRay system introduced in works of (Murphy et al., 2020), trained and validated on a dataset of 22,184 images, can perform COVID-19 detection on posteroanterior chest radiographs with averaged accuracy of 81%. The deep learning model developed in works of (Apostolopoulos and Mpesiana, 2020), trained on 1,428 CXR images, can achieve diagnostic accuracy of 94% on an imbalanced testing set. The patch-based network developed by (Oh et al., 2020) can perform 5-classes diagnosis (normal, bacterial, tuberculosis, viral and COVID-19) of the patients with averaged accuracy of 88.9% and providing interpretable saliency maps.

**1.2 Content-based Image Retrieval (CBIR) in medical image analysis**

In addition to direct diagnosis and lesion detection, there exists another commonly adopted scheme for analyzing medical images which is the content-based image retrieval (CBIR) system. Based on the idea of using image itself to perform query on a large database of images, rather than query by keyword or database structure (Mohd Zin et al., 2018), CBIR has also been widely investigated for its potential in clinical applications, such as content-based access for pathology images where pathologists can reach diagnosis by searching reference specimen slides from the existing database; as well as radiologists' reading of digital mammography where the mammogram retrieval system can provide them with intuitive visual aids for easier diagnosis (Müller et al., 2004; Müller and Unay, 2017). We thus hypothesize that a CBIR system, which can achieve near real-time medical image retrieval from massive and multi-site database for both physician/radiologist examination and computer-aided diagnosis, could be very helpful in dealing with COVID-19 pandemic. CBIR system can provide visually and semantically relevant images from a database with labels matching the query image. Thus, the label or diagnosis of the matched image can provide a clue for the queried image. The key component of a CBIR system is the embedding of images i.e. transformation of images from native (Euclidean) domain to a more representative, lower-dimension manifold, as effective image representation can enable more accurate and faster retrieval. Various image embedding methodologies specifically tailored to biomedical images have been proposed, including kernel methods such as hashing (Zhang et al., 2014), hand-crafted image filters such as filter banks (Foran et al., 2011) and SIFT (Kumar et al., 2016). Recent advancement of deep learning has also inspired CBIR systems developed based on deep neural networks (Wan et al., 2014), such as CNN for classification (Qayyum et al., 2017) and deep autoencoder (Çamlica et al., 2015) which has shown superior performance than other methods. However, current deep learning-based schemes of directly learning image representations (i.e. embeddings) based on the relationship between image features and image labels might not be the optimized approach for image retrieval task. As pointed out in (Khosla et al., 2020), comparing with cross-entropy loss which is widely adopted in current deep learning methods, pair-wise contrastive loss can be more effective in leveraging label information. Thus, in recent years, metric learning based CBIR systems for analyzing histopathological images have been developed (Yang et al., 2019, 2020). Traditional (non-deep learning) metric learning methods have also been proposed for analyzing CT (Wei et al., 2017) and magnetic resonance imaging (MRI) images (Cheng et al., 2016). To the best of our knowledge, there is no such metric learning studies for CXR images in a clinical setting.

To this end, we propose a deep learning-based CBIR system for analyzing chest radiographs, specifically for images from potential COVID-19 patients. The core algorithm of the proposed model is deep metric learning with multi-similarity loss (Wang et al., 2019) and hard-mining sampling strategy to learn a deep neural network that embeds the CXR images into a low-dimensional feature space. The embedding module has the backbone network structure of Resnet-50 (He et al., 2016). In addition, the proposed CBIR model features an attention branch using spatial attention mechanism to extract localized embeddings and provide local visualization (i.e. attention map) of the disease labels, in order to provide visual guidance to the readers and improve model performance. This design allows us to ensure both content- and semantic-similarity between the query images and the returned images.

The model is trained and validated on a multi-site COVID-19 dataset, consisting of totally 18,055 CXR images from three sources: the public open benchmark dataset COVIDx (Wang et al., 2020b), 5 hospitals from the Partners HealthCare system in MA, U.S., and 4 hospitals in Daegu, South Korea. Performance of the model is evaluated by its capability of retrieving the correct images and diagnosing the correct disease types. The proposed model is further evaluated by transferring it to a different task, where it is utilized to extract informative features from new, independently collected CXR images. Extracted features are then combined with the electronic health record (EHR) features to predict the need of intervention within 72 hours, which serves as a clinical decision support tool for COVID-19 management in emergency department.

Key contributions of this work are summarized as follows: 1) we develop a CBIR system that includes a novel embedding model with spatial attention mechanism which is trained with adjusted multi-similarity loss and hard-mining sampling strategy; 2) in both image retrieval and diagnosis tasks, the model achieves state-of-the-art performance, and shows superior performance than the Resnet-50 network which is a widely-applied method in medical image analysis; 3) the model shows high accuracy in prognosis task, and demonstrate its potential clinical values for many tasks in clinical decision support.

## 2. MATERIALS AND METHODS

### 2.1 Overview

In the workflow of our proposed CBIR system, for an incoming query CXR image, we will first extract its low-dimensional feature embedding using a deep neural network, which is trained using deep metric learning. After that, top-*k* images which are closest to the query image in the embedding space will be retrieved and displayed together with associated electronic health record (EHR). COVID-19 diagnosis of the query image can be then inferred by labels from the retrieved images. Embeddings of CXR images can be also used for other purposes such as clinical decision support. An overview of the model pipeline is illustrated in Fig. 1, details of each step especially notations for network structures can be found in section 2.3 and 2.4.

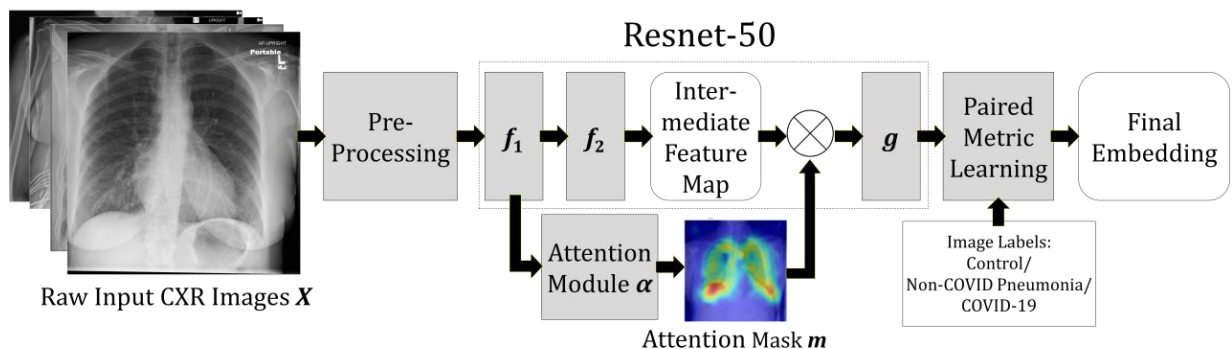

**Figure. 1** Computational pipeline of CXR image retrieval model in a COVID-19 diagnosis context.

## 2.2 Multi-site Data Collection and Description

In this study, we collected CXR images from 9 hospitals of 2 countries (5 hospitals from Partners HealthCare system in U.S, 4 hospitals from South Korea), and combined them with the public COVIDx dataset to form a multi-site dataset for training and validation. In all the three data sites, CXR images other than in the anterior-posterior (AP) or posterior- anterior PA view (e.g. in lateral view), or images with significant distortion because of on-board postprocessing (e.g. strong edge-enhancement), are excluded. Descriptions of the three data sites can be found below. It should be noted that the definition of "control" in this study includes patient with no diagnosed pneumonia nor positive PCR test results. We specifically include the type of "non-COVID pneumonia", which can be caused by a wide spectrum of reasons including bacteria, virus and fungi into this study, because it leads to similar patterns on CXR images with COVID-19, e.g. both demonstrate ground glass opacities and consolidation (Jacobi et al., 2020). In addition to the non-COVID pneumonia images in COVIDx dataset, CXR images from totally 212 patients with diagnosis of non-COVID pneumonia admitted to Partners HealthCare system during the study period were collected and included in the dataset. A brief summary and basic demographic information of this multi-site dataset can be found in Table 1.

**Table 1.** Number of hospitals and number of images involved in each data site, with break-down of patient types, average age and gender ratio.

|  | Number of Hospitals | Total Images | Control | Non-COVID Pneumonia | COVID-19 | Age | Gender |
|---|---|---|---|---|---|---|---|
| **COVIDx** | N/A | 13,970 | 8,066 | 5,551 | 353 | N/A | N/A |
| **Partners** | 5 | 823 | 107 | 212 | 504 | 58.03 | 56.6% Male |
| **Korean** | 4 | 3,262 | N/A | N/A | 3,262 | 57.31 | 35.8% Male |

Data site 1 "COVIDx": This public benchmark dataset is introduced in works of (Wang et al., 2020b), where CXR images were collected and modified from five open access data repositories. In the COVIDx dataset, we use 353 COVID-19 images (labeled as "COVID-19"), 5,551 images with non-COVID19 pneumonia (labeled as "non-COVID pneumonia"), and 8,066 images from controls.

Data site 2 "Partners": CXR scans of 5 hospitals within the Partners HealthCare system, including Massachusetts General Hospital (MGH), Brigham and Women's Hospital (BWH), Newton-Wellesley Hospital (NWH), Martha's Vineyard Hospital (MVH) and Nantucket Cottage Hospital (NCH) were collected. Patients who have received CXR imaging and had COVID-19 PCR testing in the emergency department from Dec 1$^{st}$, 2019 through March 29$^{th}$, 2020 were included, consisting of 107 CXR images from controls, 212 from non-COVID pneumonia, and 504 from COVID-19 patients.

Data site 3 "Korean": CXR scans from 4 hospitals in Daegu, South Korea, including hospitals affiliated to Kyungpook National University, Yeungnam University College of Medicine, Keimyung University school of Medicine and Catholic University of Daegu School of Medicine were collected, during the period from Feb 25$^{th}$ to Apr 2$^{nd}$. These hospitals are all in Daegu, a city of 2.5 million people which has been identified as the epicenter of the South Korean COVID-19 outbreak (Shim et al., 2020). There are totally 3,262 CXR images from hospitalized COVID-19 patients in this dataset.

## 2.3 Image Preprocessing

Preprocessing steps of CXR images in this study includes anonymization, image cropping, resizing,

windowing and lung segmentation. The major reason of including lung segmentation in the preprocessing is to prohibit the model from learning to distinguish the source of the data by features such as letters put onto CXRs, since the data collected from different sites have imbalanced label distribution. The whole lung region is automatically segmented by an ensemble of five deep neural networks. These networks have the same backbone structure of EfficientNet (Tan and Le, 2019), but with different architectures and parameters. The ensemble segmentation model is trained on one MGH dataset with 100 annotated CXRs and two public datasets: the tuberculosis CXRs from Montgomery County (Jaeger et al., 2014), and the Shenzhen and JSRT (Japanese Society of Radiological Technology) CXRs (Shiraishi et al., 2000). Data augmentation techniques are employed for training the ensemble model, including horizontal flip, Gaussian noise, perspective, sharpness, blurring, random contrast, random gamma correction, random brightness, contrast limited adaptive histogram equalization, grid distortion, affine transform, and elastic transformation. Training parameters for the ensemble segmentation model are determined through grid-search on the validation dataset, which are as follows: Adam optimizer (learning rate=0.0001), epochs=200, and batch size=8. The model is validated on an independent 122 CXRs test set with manual annotation of lung by experts. It achieved Dice coefficient of 0.95 for segmentation on the test set.

## 2.4 Content-based Image Retrieval and Metric Learning Model

Denote a set of data $(x_i, y_i)$, where $x_i$ is the CXR image of one patient, $y_i$ is the label of the patient. In this work, the label is a ternary value indicating whether the patient is from control group, has non-COVID pneumonia or COVID-19. Our goal is to learn a function $f_\theta: x \to \mathbb{R}^d$ that embeds the given CXR image into a *d*-dimensional embedding feature space, which ensures: 1) semantically same images (i.e. with the same label) shall be closer in the embedded space, and vice versa; 2) patients with similar image content, especially around lesion regions related to the disease, shall be closer in the embedded space. We employ the contrast learning scheme to find such non-linear embedding, which is a deep neural network parameterized by $\theta$. It has been reported in previous literatures that learning representations by contrasting positive pairs against negative pairs can be more advantages than learning the direct mapping from data to its label for improved robustness and stability (Hadsell et al., 2006). To achieve these two goals, we adopt a metric learning scheme to train the network with paired input images as input and multi-similarity loss between the image pairs as loss. We also exploit spatial attention mechanism to focus the model on potential lesion regions. Attention mechanism allows salient features to be dynamically localized to the forefront as needed (Xu et al., 2015) and has been widely used in many applications such as image segmentation (Fu et al., 2019) and classification (Wang et al., 2017a).

### 2.4.1 Loss Function and Sampling Strategy

In this work, we use the cosine similarity $S$ between embedded features to measure the similarity between pairs of images, namely:

$$S_{i,j} = \frac{\langle f(x_i), f(x_j) \rangle}{\|f(x_i)\|_2 \|f(x_j)\|_2}, \tag{1}$$

where f is the embedding function we aim to learn. Following the common practice in metric learning, we will normalize the embeddings at the end, letting $\|f(x)\|_2 = 1$ for all $x$.

We employ the multi-similarity loss (Wang et al., 2019) for the "paired metric learning" step in Fig. 1, which has achieved state-of-the-art performance on several image retrieval benchmarks. The loss function $L$ is adjusted to our setting by:

$$L = \frac{1}{m}\sum_{i=1}^{m} \frac{1}{\alpha}\log\left[1 + \sum_{j \in P_i} e^{-\alpha(S_{i,j}-\lambda)}\right] + \frac{1}{\beta}\log\left[1 + \sum_{j \in N_i} e^{\beta(S_{i,j}-\lambda)}\right], \tag{2}$$

where $P_i$ and $N_i$ are the indices set of selected "same type" (i.e. images with the same label) and "different types" (i.e. images with different labels) pairs of samples regarding to the anchor image $x_i$, $m$ is the batch size and $\alpha$, $\beta$, $\lambda$ are hyperparameters. For each minibatch during training, we randomly select $N$ samples from each class, forming a minibatch of size $T \times N$, where $T$ is the number of classes. Every two samples in the batch can be used as a pair in the calculation of the loss function.

Training with random sampling may harm the capacity of the model and slows the convergence (Wu et al., 2017), since pair-based metric learning often generates large number of sample pairs which can include informative easy or redundant pairs. We use a hard-mining strategy to improve model performance and speedup training convergence: each "same type"/"different types" pair will be compared to the hardest pairs in the whole batch to mine the hard pairs, as performed in (Wang et al., 2019).

### 2.4.2 Spatial Attention Mechanism for Localized Feature Extraction

Spatial attention mechanism is adopted in our embedding model to obtain disease-localized embeddings of the patients and to provide interpretable output at the stage of image retrieval. Specifically, an attention module is plugged into the network in parallel with feature extraction route represented by $\alpha(\cdot)$, which generates a mask with the value from 0 to 1 and the same spatial dimension of network's intermedia feature map. The attention route in Fig. 1 illustrates how the attention module is plugged into the backbone network. Element-wise multiplication will be performed between the output attention mask and intermedia feature map of the network to obtain a localized feature map. This localized feature map is then sent to the projection head to get the final embedding. In other words, by writing the embedding function as:

$$f(x_i) = g\left(f_2(f_1(x_i))\right) \qquad (3)$$

where $f_1$ and $f_2$ are different stages of the feature extractor (i.e. convolutional layers) and $g$ is the projection head which projects the representations into lower-dimensional embedding space, shown as the corresponding lettered blocks in Fig. 1. As the embedding will be served as input to the later metric learning module, the projection aims to reduce the dimension of embedding for improved performance. The final embedding with plugged spatial attention module is:

$$\tilde{f}(x_i) = g\left(\alpha(f_1(x_i)) \odot f_2(f_1(x_i))\right). \qquad (4)$$

In Eq. 4, output of the network $f_1(x_i)$ goes through attention module $\alpha(\cdot)$ to generate the attention mask $m(x_i) = \alpha(f_1(x_i))$, which localizes the intermediate feature map (i.e. $f_2(f_1(x_i))$) in Fig. 1 before feeding into the projection head $g$. This whole embedding model will be then optimized by the metric learning scheme as introduced previously. This design is inspired by the work in (Kim et al., 2018), in which attention modules enables computer vision algorithm to attend to specific parts of the object.

### 2.4.3 Implementation Details and Source Code

We use Resnet-50 (He et al., 2016) as the backbone architecture for the feature extraction. The first stage $f_1$ is consisted of the first part of Resnet-50 until conv3_4 (the 22$^{nd}$ layer of Resnet-50), and the second stage $f_2$ is consisted of the later part of Resnet-50 until conv4_6 (the 40$^{th}$ layer of Resnet-50). The projection head $g$ includes the remaining part of the Resnet-50 and two fully connected layers that project the extracted features into a 64-dimensional embedding space. Attention module is place between conv3_4 and conv4_6 following the similar practice in the works of (Kim et al., 2018). The attention module takes the output of block 3 of Resnet-50 as input. It then generates masks of size 16*16 which is later applied to the output of block 4 of the Resnet-50. Architecture of the attention module we use consists of 3 "bottleneck" building blocks in the Resnet, followed by a Squeeze-and-Excitation layer (Hu et al., 2019), channel-wise averaging and sigmoid activation. All the CXR images are resized to 256×256 with the aspect ratio fixed for both training and testing. We randomly crop images to 256×256 during training but use the whole image during testing. We used Adam optimizer with default parameters. The learning rate is set to 3e$^{-5}$. We trained our model for 2,000 iterations with batch size $T \times N = 3 \times 16 = 48$, which is roughly equivalent to 5 epochs, using pretrained model from ImageNet (Deng et al., 2009) as initialization. Parameters in the loss function are set as $\lambda=0.5$, $\alpha=2$ and $\beta=20$, derived from grid-search. For the purpose of classification, we employ the K-nearest Neighbor (KNN) classifier (i.e. returning $k$ nearest images based on distance in the embedding space) with distance weighting (i.e. closer neighbors of a query point have larger weight). In this work we set $k=10$, that is, for each query image 10 neighbor images will be retrieved by the model, which then make the weighted majority vote to determine the label of query image. Label of the query image is then determined by the weighted majority vote from the label of returned $k$ images. The weighted voting also avoids a tie. Source code of the model, also including trained network and CXR preprocessing modules, will be published on a public repository (GitHub), available to be downloaded and used by public. Images from the COVIDx dataset used in this work will be shared along with the codes for easy replication and testing.

# 3. RESULTS

Here we present our results of CBIR-based modelling and processing of COVID-19 CXR images in three perspectives: validity of the model by its capability of performing correct image retrieval and comparison with baseline method; clinical value of the model by its multi-site diagnostic performance; and finally transferability of the model by using its embedding function for a different clinical decision support task.

## 3.1 Image Retrieval Performance and Comparison with Baseline Method

**Table 2.** Sample sizes and splitting of training/validation data of the three (COVIDx, Partners and Korean) data sites used in this work.

|  | Train | | | | Validation | | | |
| --- | --- | --- | --- | --- | --- | --- | --- | --- |
|  | Total | COVIDx | Partners | Korean | Total | COVIDx | Partners | Korean |
| **Control** | 8,064 | 7,966 | 98 | N/A | 109 | 100 | 9 | N/A |
| **Non-COVID Pneumonia** | 5,641 | 5,451 | 190 | N/A | 122 | 100 | 22 | N/A |
| **COVID-19** | 3,746 | 253 | 453 | 3040 | 373 | 100 | 51 | 222 |

The multi-site dataset is split into training and validation part according to Table 2. Patient types are varying across different data sites, so we performed the splitting to ensure that maximum number of sites are presented in both training and validation data, to remove potential site-wise bias. As there is no label of "non-COVID pneumonia" in the Partners data site (labels are determined based on PCR test), and no "control" nor "non-COVID pneumonia" in the Korean data (all COVID-19 patients), there are several "N/A (not available)" entries in Table 2.

After training the proposed model to learn the feature embeddings, we performed the image retrieval task using a neighbourhood size $k$=10 (i.e. ten images will be returned by the model for each query). Due to the space limit, we only demonstrate and analyze the results using the top 4 returned images. Sample query/return CXR images and clinical information of the returned images are visualized in Fig. 2. Because of limited space, we only show important clinical information here, including patient gender, age, Radiographic Assessment of Lung Oedema (RALE) score (Warren et al., 2018), SpO2 (oxygen saturation), WBC (white blood cell count), admission to ICU (intensive care unit). RALE is originally designed for evaluating CXRs of acute respiratory distress syndrome (ARDS). As COVID-19 is similar and will potentially lead to ARDS, we are using RALE here to roughly assign COVID-19 images to "mild" cases as in Fig. 2(a), and "severe" cases as in Fig. 2(b). It should be noted that RALE scores of each CXR image are manually assessed by two senior radiologists in Partners healthcare group, thus they are only available in the "Partners" and "Korean" data for the purpose of validating our results. In the future, an AI based model will be used to automatically estimate RALE score in the EHR system so that the score also appears in the retrieved clinical information. Also, these is no clinical information available in the public "COVIDx" data site. From the returned CXR images it can be found that: 1) CXR images from different data sites with the query image but of the same label can be correctly retrieved, indicating that there is little site-wise bias of the learned embedding; 2) The model can handle image with heterogenous patient characteristics e.g. varying sizes of the lung and varying locations of lesion regions, as well as heterogenous imaging conditions; and 3) We can observe a strong similarity of patient's severity among the retrieved images, as shown in panel (a) and (b) in Fig. 2. Specifically, both the RALE score and patient's admission to ICU indicate that the four returned images in Fig. 2(b) are consistently more severe than the returned images in Fig. 2(a). As the RALE score of the query images in Fig. 2(a) and (b) are 2 and 34 respectively, we find the severity of returned images are also related to the patient's condition of the query image. Considering the fact that the model is trained without patient's severity of disease (i.e. only based on three types of image labels), its ability in retrieving severity-associated images shows that it can correctly extract CXR features that are sensitivity to COVID-19's disease progression.

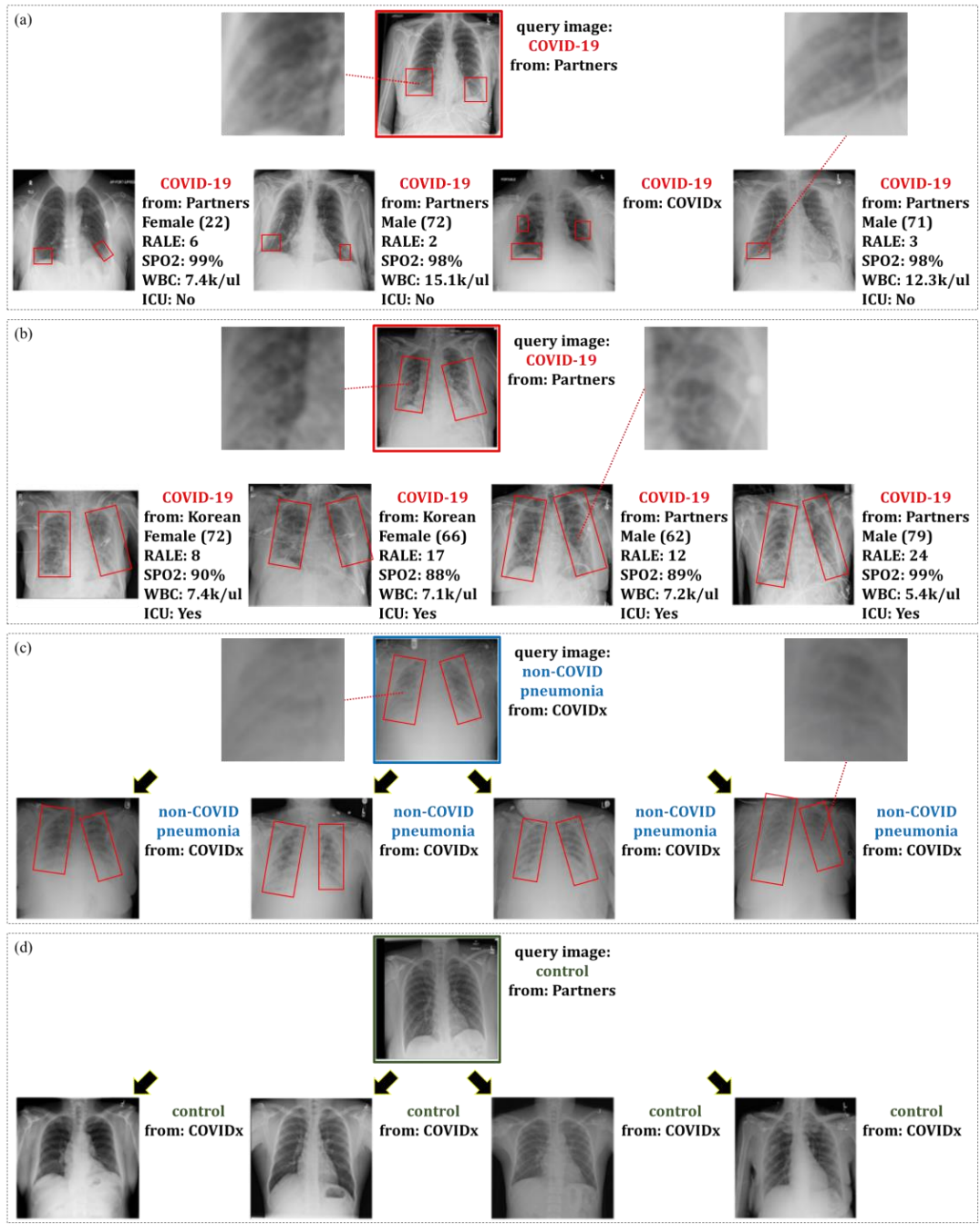

**Figure 2.** (a) Sample visualizations of the returned CXR images by the proposed model with query CXR image from a mild COVID-19 patient. Possible lesion regions are marked by red bounding boxes, with zooming in to detailed textures in the lesion region. (b) With query CXR image from a severe COVID-19 patient. (c) With query CXR image from non-COVID pneumonia patient, note that only COVIDx dataset contains this type of images. (d) With query CXR image from control, note that about 99% of the controls are from COVIDx dataset.

In order to investigate the effectiveness of the attention module as introduced in section 2.4, attention maps generated by the proposed model of three samples images are visualized in Fig. 3. We select images from COVID-19 patients with different RALE score indicating their severity of disease. In Fig. 3, for the image to the left (RALE score=2), its opacities are mainly on bilateral lower quadrants with extent <25%, where its attention map also shows that the majority of model attentions are at both lower lung regions. For the image at middle (RALE score=8), it has opacities with moderate density occupy 25-50% of bilateral lower quadrants. Its attention map is focused on the same lower quadrants of both left and right lung with higher coverage. For the image to the right (RALE score=25), there are moderate to dense opacities in all four quadrants of the lung: extent of consolidation is 25-50% in right lung and upper quadrant of left lung, 50-75% in lower left quadrant. Attention map of this image covers all area of the lung, especially focuses on the right lower quadrant. Such correspondence between human observation study through RALE score and attention maps shows that the attention mechanism employed by the proposed model can correctly localize potential lesion regions of the lung. Thus, the attention module can offer improved discriminability for the feature embeddings learnt by the model, by only keeping the most disease-related features.

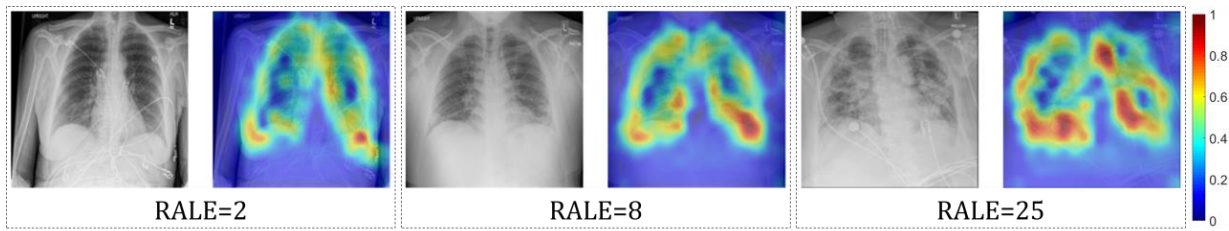

**Figure 3**. Visualization of CXR images and the corresponding attention maps from COVID-19 patients with different RALE score, which indicates their severity of disease.

To quantitatively evaluate the performance of the proposed model on this query by example task, we calculate the averaged recall rate of the *k* returned images over all test samples. A query sample is defined as "successfully recalled" if at least one image in the *k* returned images has the same label of query image. For reference, as the dataset involved in this work has a single label of three classes, a random retrieving model will have averaged recall rate of 33.3% when *k*=1, 55.6% when *k*=2, 81.0% when *k*=4, and 95% when $k$=10 on a balanced dataset. Recall rates of the proposed model with different parameter $k$ are listed in Table 3 (left).

**Table 3.** Model performance comparison between the proposed and baseline model, evaluated by averaged recall rate across all validation samples under different parameter $k$.

|  | Proposed System | | | | Basline (Resnet-50) | | | |
| --- | --- | --- | --- | --- | --- | --- | --- | --- |
|  | $k$=1 | $k$=2 | $k$=4 | $k$=10 | $k$=1 | $k$=2 | $k$=4 | $k$=10 |
| **Control** | 66.1% | 81.7% | 84.4% | 93.6% | 74.3% | 89.0% | 95.4% | 97.2% |
| **Non-COVID Pneumonia*** | 87.7% | 91.8% | 91.8% | 94.3% | 82.8% | 87.7% | 90.2% | 93.4% |
| **COVID-19** | 83.6% | 87.9% | 90.1% | 92.5% | 80.4% | 86.3% | 89.8% | 92.5% |

For comparison, the baseline image retrieval model was developed based on a raw Resnet-50 network following traditional classification scheme. The network was trained using CXR images as input and the ternary image labels as output, with cross-entropy loss. We then extract the intermediate output from the last global average pooling layer and use it as feature embeddings for the input images. The same cosine similarity in Eq. 1 is used to measure the similarity between embeddings, which is then used for image retrieval. Pipeline of this baseline image retrieval model based on Resnet-50 is illustrated in the top panel of Fig. 4, with comparison of example retrieved images in the bottom panel of Fig. 4. As shown in the example retrieval task, our proposed model can retrieve more similar images with the correct labels, comparing with the baseline model. Performance of the baseline model for the same image retrieval task is listed in Table 3 (right). The quantitative evaluation in Table 3 shows that our proposed model achieves higher recall rate in the task of retrieving non-COVID pneumonia and COVID-19 CXR images, which is a more important task for COVID-19 screening and resource management. For the task of retrieving normal control images, the proposed model performs slightly worse than the baseline model. Investigation into model outputs reveal that the baseline model is more likely to retrieve images from the same dataset to the query image. Because the majority of normal control images come from a single (COVIDx) dataset, the baseline model can achieve better recall rate. That is also the reason why the proposed model has better performance for non-COVID pneumonia and COVID patients.

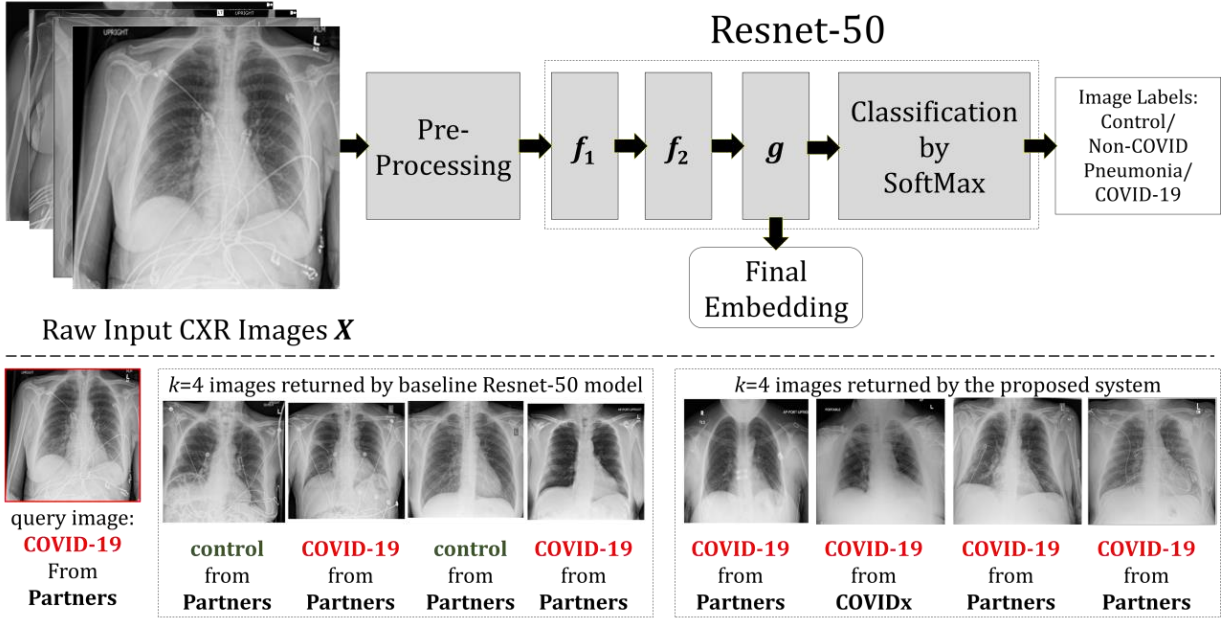

**Figure 4.** Top panel: pipeline of the image retrieval process implemented by the baseline direct classification network (raw Resnet-50). Bottom panel: comparison of retrieved top four images between the baseline model and proposed model, using the same sample query image (COVID-19 from Partners). Image retrieved by the proposed model (the same as in Fig. 2) are also listed here for reference.

### 3.2 Classifying Control, Non-COVID Pneumonia and COVID-19 Patients

We further evaluate the potential clinical value of the proposed model by its diagnostic performance. In the proposed model, label of the query image was determined by the majority vote of labels from the returned neighbour images. Diagnosis results are listed in Table 4. Sensitivities and positive predictive values (PPVS) of non-COVID pneumonia and control are not available to the Partners and Korean dataset as there are no images with the corresponding labels in these two sites. Overall, the proposed model can achieve >83% accuracy in performing COVID-19 diagnosis. Most notably, it achieves very high sensitivity for non-COVID pneumonia and COVID-19 (>85%), indicating that the model can potentially serve as a screening and prioritization tool right after the chest radiography scan is performed. We also evaluate the performance of baseline method, the raw Resnet-50 network described in section 3.1, by applying it on the validation data. Its performance is listed to the right panel of Table 4. As the raw Resnet-50 network is trained for the very purpose of classifying images by their labels, it is expected that the baseline method can achieve good performance on this diagnosis task. However, comparison between the two models shows that the proposed model outperforms baseline Resnet-50 model in overall performance in all three types (control, non-COVID pneumonia and COVID-19) of images. While in the COVIDx dataset the two models have very similar performance, the proposed model achieved better accuracy in classifying non-COVID pneumonia patients in Partners dataset. Such task is specifically difficult as data from non-COVID pneumonia patients were acquired together with COVID patients using the same machine and protocols, thus they are more homogenous and difficult to separate. On the contrary, non-COVID pneumonia population in COVIDx dataset are acquired from separate sources than COVID patients. Also, it should be noted that while the images in Korean data (totally 222 images, all COVID-19) for validation can be easily diagnosed by both models, overall diagnosis of COVID-19 only by CXR images is still a difficult task, which has also been recognized by radiologists (Murphy et al., 2020). In summary, the result indicates that the proposed metric learning scheme has a higher level of capability to learn a label-discriminative embedding from the input images.

**Table 4.** Model performance evaluated by the averaged accuracy, sensitivity and PPV for each type in the validation dataset. Left panel: performance of the proposed model. Right panel: performance of the baseline Resnet-50 model. Better performance between the two models are highlighted by bold text.

|  | Proposed System | | | | Baseline (Resnet-50) | | | |
| --- | --- | --- | --- | --- | --- | --- | --- | --- |
|  | Overall | COVIDx | Partners | Korean | Overall | COVIDx | Partners | Korean |
| Averaged Accuracy | **83.9%** | **76.7%** | **72.0%** | **98.2%** | 81.5% | 75.3% | 61.0% | 97.3% |
| Sensitivity : Control | 74.3% | 75.0% | 66.7% | N/A | **76.1%** | **77.0%** | 66.7% | N/A |
| PPV: Control | **79.4%** | **90.4%** | 31.6% | N/A | 74.8% | 86.5% | 31.6% | N/A |
| Sensitivity:Non-COVID Pneumonia | **89.3%** | 93.0% | **72.7%** | N/A | 82.8% | **95.0%** | 27.3% | N/A |
| PPV: Non-COVID Pneumonia | **64.5%** | 62.8% | **94.1%** | N/A | 61.6% | **63.3%** | 54.5% | N/A |
| Sensitivity: COVID-19 | **85.0%** | **62.0%** | 72.5% | **98.2%** | 82.6% | 54.0% | **74.5%** | 97.3% |
| PPV: COVID-19 | **95.2%** | **89.9%** | **80.4%** | 100.0% | 93.6% | 88.5% | 73.1% | 100.0% |

## 3.3 Ablation Study

### 3.3.1 Effect of Spatial Attention Mechanism

As described in Section 2.4.2, spatial attention mechanism is utilized in this work to focus the image embedding towards disease-specific regions. In order to investigate the effectiveness of attention mechanism, we implement the CBIR-based model using the identical model structure and hyperparameters and train it on the same dataset, but without the attention module $\alpha(\cdot)$ and the corresponding attention mask $m(x_i)$. Comparison between the model with and without attention mechanism on the testing dataset shows that attention mechanism can lead to a near 1% performance improvement in classification task (accuracy of 82.95% without, 83.94% with attention module). We also investigate how the cross-entropy based image retrieval model (i.e. baseline model) can benefit from attention mechanism by similarly implement and train a Resnet-50 network without attention module. Results show that attention module can contribute to near 5% performance improvement to the baseline model (classification accuracy of 76.99% without, 81.46% with attention module). Finally, it is found that attention can improve the recall rate, as described in section 3.1. Using $k$=4, the proposed model can achieve recall rates of 84.4%, 91.8% and 90.1% for control, non-COVID pneumonia and COVID, respectively (listed in Table 3), while the corresponding recall rates of the model without attention are 67.0%, 89.3% and 91.4%.

### 3.3.2 Effect of Different Contrastive Loss

We utilize the multi-similarity loss (Wang et al., 2019) in this work for training the image retrieval network. As there exists other type of contrastive loss functions, here we investigate the performance of an alternative model using the Noise-Contrastive Estimation (InfoNCE) loss (Oord et al., 2018), which has been widely applied in both self-supervised and supervised contrastive learning. InfoNCE loss is based on optimizing categorical cross-entropy for classifying one positive sample from $N$ random samples consisting of $N$-1 negative samples, where these $N$ samples were sampled from a proposal noise distribution. Comparison between the proposed model and the model using InfoNCE loss (with everything else remained the same) show similar performance (accuracy of 83.94% by proposed model, 82.78% by InfoNCE).

### 3.3.3 Ablation Study of Hyperparameter Setting for KNN Classifier

As the CBIR-based model relies on KNN to obtain labels for the query images, we investigate how the number of returned nearest neighbors to be considered in making the weighted majority vote (i.e. value of *k* for KNN) can affect model performance. By trying different values of *k* from 1 to 30, it is found that classification accuracy is stable when the *k* is within a reasonable range (5~20), as illustrated in Fig. 5(a). This is mainly because we weight the returned neighbors based on their distance to the query image for making the majority vote, thus the increased neighbors because of a larger *k* will have reduced impact on voting results. Thus, we use *k*=10 for the proposed model based on empirical experiment and efficiency.

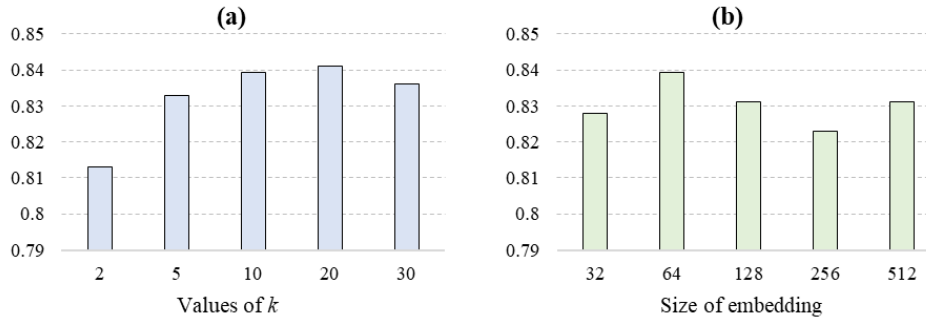

**Figure 5.** (a) Model performance (as measured in classification accuracy) with different value of *k* for KNN classifier. (b) Model performance with different size of image embedding, which is served as input to the metric learning module.

### 3.3.4 Ablation Study of Hyperparameter Setting for Image Embedding

As introduced in section 2.4.2 and 2.4.3, we use the projection head *g* to project the learned image representations into lower-dimensional embedding space. As the feature extracted by Resnet-50 has dimension of 2048, the projection head will project this 2048-D feature into a smaller size, where we have investigated different possible size ranging from 32 to 512. Model performance by using different embedding size are illustrated in Fig. 5(b). As there exists a trade-off between image information preserved after embedding (which prefers a larger embedding space) and the dimensionality problem for later metric learning (which prefers a smaller embedding space), the optimized size for embedding space is highly relied on the later task and data distribution thus can only be determined empirically. In the current model setting we use the embedding size of 64, based on the consideration of both model performance and efficiency.

### 3.4 Transferring Embedded Image Features for Clinical Use

As introduced in the methodology development section, the proposed model is developed with the aim of learning both content- and semantic-rich embeddings from the input images. Thus, after training, the model can be also used as an effective image feature extraction tool for other tasks based on the learned embeddings. In order to test the feasibility of the proposed model on such premise, we employ the pre-trained model on a new task of clinical decision making. The task is part of our Partners healthcare institution's goal of predicting the emergency department (ED) COVID-19 patient's risk of receiving intervention (e.g. receiving oxygen therapy or under mechanical ventilator) within 72 hours. Such prediction is strongly correlated to prognosis and is vital for the early response to patients and management of resources, which can be beneficial for both patients and hospital. On one hand, intervention measures especially ventilators have been recommended as a crucial for the countering the hypoxia of COVID-19 patients (Orser, 2020), where timely application of intervention has been considered as an important factor to patient's prognosis (Meng et al., 2020). On the other hand, effective resource allocation of oxygen supplement and mechanical ventilator has become a major challenge during COVID-19 epidemics, thus the knowledge of equipment needs in advance will be helpful for the hospitals, especially in emergency department.

Electronic health record (EHR) data and CXR images were collected from 1,589 COVID-19 PCR test positive patients who has been admitted to emergency department of the hospitals affiliated with Partners group before April 28$^{th}$, 2020. In total 17 EHR-derived features were used in this study after a feature selection using random forest. These features include patient's demographic information (e.g. age), vitals (e.g. temperature, blood pressure, temperature, respiratory rate, oxygen saturation, etc.), and basic lab tests (e.g. glomerular filtration rate, white blood cell, etc.). 2048-dimensional CXR-derived image features (i.e. features extracted by Resnet-50 backbone with attention, before processed by projection head *g*) were extracted using the proposed model, which has been pre-trained as in section 3.1 without any further calibration to the data in this task. Types of intervention the patients have received for breathing, including high flow oxygen through nasal cannula, non-invasive ventilation through face mask and mechanical ventilators in 72 hours, were recorded as the prediction target.

We then trained 3 binary classifiers to predict whether the patient will be receiving any types of interventions. The first classifier uses only EHR-derived features as input for the prediction, the second classifier uses only CXR-derived features as extracted by the proposed model as input, and the third classifier uses the combined CXR+EHR features as input. We tried different classification methods including logistic regression, SVM, random forest and the Deep & Cross network (Wang et al., 2017b) with different hyperparameter settings for this experiment, and only reported the results with best performance. Specifically, for Deep & Cross network we employ a Multilayer Perceptron (MLP) with two 128-dimensional fully connected layers, and a two-layers Cross Net. The network is trained with Adam optimizer using lr=0.0001 of 10 epochs. For random forest model we use max depth of 5, with 50 number of estimators. For classification using only EHR-derived features or CXR-derived features, we used the random forest classifier. For the classification using combined features, we used the Deep & Cross network. Prediction models were evaluated by their receiver operating characteristic (ROC) using 5-folds cross-validation, as shown in Fig. 6. The averaged area under curve (AUC) of the CXR-only prediction model is 0.831, and the AUC of CXR-EHR combined prediction is 0.913. This result validates the prediction model's feasibility in providing estimation of patient's condition upon his/her admission, either using only CXR scan or using combined CXR-EHR features. As no calibration to the data is needed, our proposed image retrieval model has shown its capability of image features extraction, which can be universally applied to a wide spectrum of similar tasks in clinical decision support. In other words, any CXRs collected in a COVID-19 related task can be potentially processed by the proposed model to obtain their feature embeddings. On the other hand, we see that adding CXR features into the prediction can improve its performance especially for robustness: prediction using only EHR-derived features has an averaged AUC of 0.887 yet with much higher standard deviation of 0.025, indicating its relative instability of prediction.

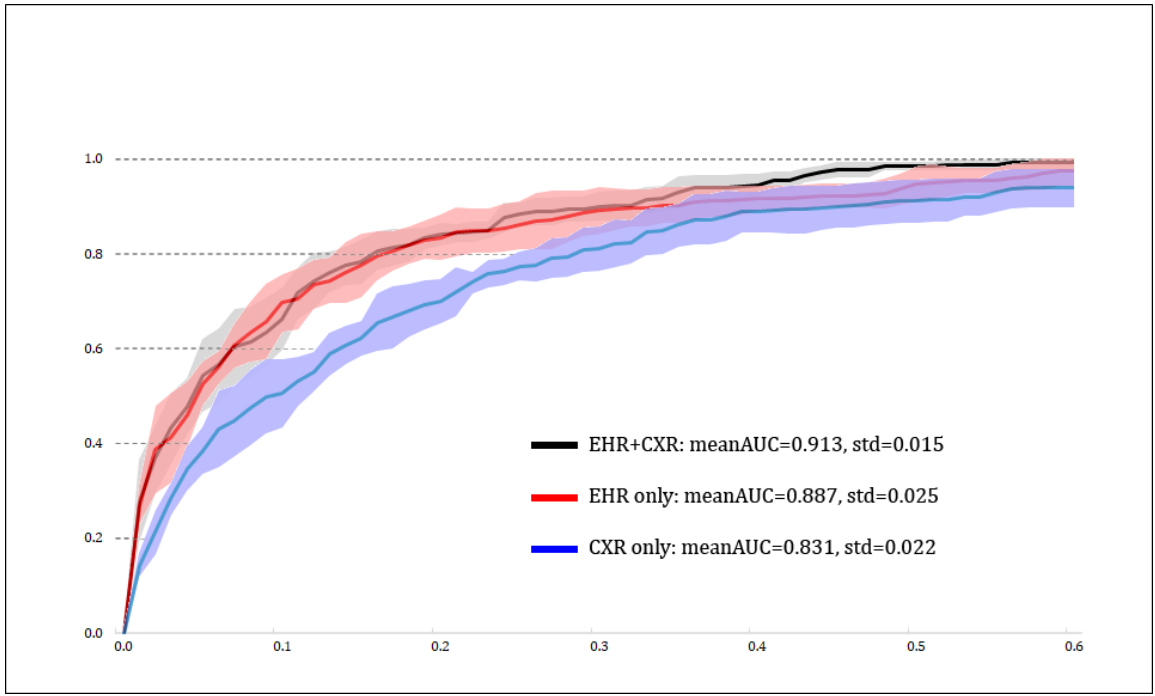

**Figure 6.** ROC curve of the 72-hours patient intervention prediction model using combined features (black), CXR-derived (blue) and EHR-derived (red) features as input. Mean ROC across the 5 cross-validation is illustrated as solid curve, ±1 standard deviation is illustrated as areas around the mean curve.

## 4. DISCUSSION AND CONCLUSION

In this work we proposed a metric learning based CBIR model for analyzing chest radiograph images. Based on the experiments, we show that the proposed model can handle a variety of CXR-related clinical problems in COVID-19, including but not limited to CXR image feature extraction and image retrieval, diagnosis, and clinical decision support. Comparison with traditional classification-based deep learning method shows that the metric learning scheme adopted in this work can help improving effectiveness of image retrieval and diagnosis while at the same time providing rich insights into the analysis procedure, thanks to the model's capability in learning both semantic and content discriminative features from input images. In addition, the clinical information returned by the retrieval model, as illustrated in Fig. 2, can provide reference for the radiologists and physicians in determining the query patient's condition to assist decision making. Such capability of linking image and clinical information through content-based retrieval will be extremely helpful for the radiologists and physicians in facing the potential threat of a COVID-19 resurgence.

The superior performance of the proposed model in retrieving images for radiologists and physicians, and its value in diagnosis/prognosis has motivated our Partners healthcare consortium to start deploying the model into clinical workflow and integrating it in the EHR system (e.g. EPIC system as used in Partners healthcare). Significant amount of engineering and integration work has been done in this effort. In addition to data routing, series selection and interface development for the system integration, we have been specifically working on: 1) improving the model for a more comprehensive query strategy i.e. incorporating keyword- and clause-based query; 2) establishment of a standardized definition of COVID-19 clinically relevant patient features, which will be identified from patient's EHR data, extracted and routed by the system, and displayed to the human readers along with returned images; 3) the development of institutional-level COVID-19 data warehouse to support large-scale, holistic coverage for COVID-19 data collection within the Partners healthcare system.

In the current study, the proposed model is applied on a single-label, three-classes task. As the multi-similarity loss enforced during metric learning process is intrinsically designed for learning from multi-labeled data, the model can be easily adapted to more challenging, multi-label tasks such as identifying lung-related comorbidities in COVID-19 patients. As comorbidities such as chronic obstructive pulmonary disease (COPD) and emphysema can interfere with the severity assessment of COVID-19, correct identification of those conditions during image retrieval will be very important and useful. Towards this purpose, richer semantic information (i.e. more disease labels) and data collection from a larger population will be included in our future study. Further, we are extending the current patient types (control, non-COVID pneumonia, COVID-19) into a wider range of definition. By incorporating severity level of COVID-19 as reported by the physicians into analysis, we can develop an improved version of the model with capability of discriminating and predicting patient severity.

Another major challenge of the content-based image retrieval is the definition of "similarity". As discussed in (Smeulders et al., 2000), there exists "semantic gap" between information extracted by computer algorithms from an image and perception of the same image by human observer. Such gap is more prominent in medical domain, as semantic disease-related features are usually localized with very specific texture definition, while visual perception of the image is more focused on global shape and position of the lung in CXR images. Thus, it will be difficult to interpret image retrieving results by the radiologists, especially when multiple labels are involved in the reading. To address this challenge, we are working on the development of a more user-friendly system, in which human readers can obtain different outputs by adjusting a hyperparameter to control the balance between semantic and visual similarities.